\documentclass[journal]{IEEEtran}
\usepackage{xr-hyper} 
\usepackage{hyperref} 
\usepackage[acronym,nogroupskip,style=super]{glossaries}
\usepackage{enumitem}    
\usepackage{hyperref}
\newacronym{iot}{IoT}{Internet of Things} 
\usepackage{cite}
\usepackage{amsmath,amssymb,amsfonts}
\usepackage{graphicx}
\usepackage{textcomp}
\usepackage{bmpsize}
\usepackage[dvipsnames]{xcolor}
\usepackage{lipsum}
\usepackage[font=footnotesize,labelfont=bf]{caption}
\usepackage[export]{adjustbox}
\usepackage{pifont}
\usepackage{multirow}
\usepackage{tabularx}
\usepackage{algorithmicx}
\usepackage[noend]{algpseudocode}
\usepackage[T1]{fontenc}
\usepackage{algorithm} 
\usepackage{subfig}	
\usepackage{soul}
\usepackage{booktabs}
\usepackage{multirow}
\usepackage{rotating}
\usepackage{booktabs}
\usepackage{siunitx,etoolbox}
\usepackage{color, colortbl}
\usepackage{comment}
\floatname{algorithm}{SM-Algorithm}
\newcommand{\beginsupplement}{%
	\setcounter{table}{0}
	\setcounter{page}{1}
	\setcounter{equation}{0}
	\setcounter{section}{0}

	\setcounter{figure}{0}
	\renewcommand{\thefigure}{S\arabic{figure}}%
}
\usepackage{changepage}     
\begin{document}
\title{IoTDevID: A Behavior-Based Device Identification Method for the IoT}
\author{Kahraman~Kostas,
		Mike~Just,
		and~Michael~A.~Lones,~\IEEEmembership{Senior Member,~IEEE}
\thanks{All authors  are with the Department of Computer Science, Heriot-Watt University, Edinburgh EH14 4AS, UK,  e-mail: \{kk97, m.just, m.lones\}@hw.ac.uk}
\thanks{KK supported by Republic of Turkey - Ministry of National Education.}}
\markboth{IEEE INTERNET OF THINGS JOURNAL,~Vol.~XX, No.~X, XXX~2022}
\makeatletter
\def\ps@IEEEtitlepagestyle{%
  \def\@oddfoot{\mycopyrightnotice}%
  \def\@oddhead{\hbox{}\@IEEEheaderstyle\leftmark\hfil\thepage}\relax
  \def\@evenhead{\@IEEEheaderstyle\thepage\hfil\leftmark\hbox{}}\relax
  \def\@evenfoot{}%
}
\def\mycopyrightnotice{%
  \begin{minipage}{\textwidth}
  \centering \scriptsize
  Copyright~\copyright~2022 IEEE. Personal use of this material is permitted. However, permission to use this material for any other purposes must be obtained from the IEEE by sending a request to pubs-permissions@ieee.org.
  \end{minipage}
}
\makeatother
\maketitle

\begin{abstract}
Device identification is one way to secure a network of IoT devices, whereby devices identified as suspicious can subsequently be isolated from a network. In this study, we present a machine learning-based method, \textit{IoTDevID}, that recognizes devices through characteristics of their network packets. As a result of using a rigorous feature analysis and selection process, our study offers a generalizable and realistic approach to modelling device behavior, achieving high predictive accuracy across two public datasets. The model's underlying feature set is shown to be more predictive than existing feature sets used for device identification, and is shown to generalize to data unseen during the feature selection process.
Unlike most existing approaches to IoT device identification, IoTDevID is able to detect devices using non-IP and low-energy protocols.
\end{abstract}

\section{Introduction}
The Internet of Things (IoT) is a collection of objects with embedded systems that can communicate with each other over a wired or wireless network. 
The ``things'' can be any physical items 
that we have 
now or will use in the future~\cite{hussain2020machine}. IoT contributes 
to human life in many critical areas such as smart homes/cities, 
retail, 
healthcare, transportation, agriculture, military, 
and manufacturing~\cite{kouicem2018internet}. 
By 2026, the number of IoT devices in the world is expected to reach 80 billion~\cite{rosenmike}, with a total market of 1.1 trillion USD~\cite{fortunebusinessinsights}. 

Securing IoT devices with traditional security solutions can be challenging due to limited device resources, such as processor, battery and bandwidth~\cite{zarpelao2017survey}.
Further challenges arise due to device heterogeneity~\cite{hussain2020machine}.
For example, a device can have  many sensors (temperature, humidity, motion, light, etc.), and the channels it uses to communicate with other devices may need very different requirements. Although some research focuses on specialist areas such as home appliances~\cite{alrawi2019sok} or smart cities~\cite{ozay2015machine}, many devices have very different characteristics even under this classification. For example, baby monitors and smart kettles are both classified under home appliances\cite{alrawi2019sok}. Further,
IoT security has the potential to be neglected by users,
primarily due to the lack of a familiar interface~\cite{alrawi2019sok}.     However, since today's IoT devices are widespread and diverse, users can not be expected to recognise and understand the security implications of all of them. Hence, automatic identification  of devices in the network is vital in dealing with 
potential security
problems. A successful device identification (DI) system can detect devices in a network, so necessary measures such as updating, limiting, or isolating devices with security vulnerabilities can be taken. In this way, potential security problems caused by IoT devices on the network can be prevented before they arise.

In this paper, we introduce a new IoT DI method that models the behavior of the network packets communicated by the devices. 
With IoTDevID, devices are identified by their behavior, 
a critical first step to the subsequent detection of anomalous behavior. 
It classifies device behaviors at the individual packet level using generalizable features. In this way, it can detect all kinds of devices, while offering high detection success with its aggregation method. This work offers the following contributions and differences from previous work:
\begin{enumerate}
  \item Our method does not  strictly depend on identifying features such as the MAC and IP address in the feature extraction step. This gives it wider applicability than previous approaches that depended on this information, such as~\cite{miettinen2017iot,hamad2019iot,bezawada2018behavioral}. Notably, it allows identification of devices that use non-IP and low-energy protocols such as Bluetooth, ZigBee, or ZWave.
  \item Our DI model uses a feature set that summarizes the network behavior of IoT devices. We carried out a systematic study of packet-level features, using various feature selection techniques to build a feature set that improves on previous approaches. Unlike previous studies~\cite{aksoy2019automated,hamad2019iot,sivanathan2018classifying}, we were careful to identify and remove features (such as session IDs and port numbers) which do not generalise between datasets.
  \item Our approach was developed using a rigorous experimental process. We took steps to prevent information leaking from the test set into training, and used a second, independent, dataset to demonstrate the generality of our approach. Consequently, we believe that our results are more realistic and generalizable than much of the previously published work in this area.

\end{enumerate}
 
As a consequence, our DI method is more suited to real world IoT network environments than previous approaches, both because it supports a wider range of environments, but also because we have more confidence that the approach generalises to environments that were unseen during  development. To promote transparency and repeatability, we have made our dataset, scripts, and supplementary material public\footnote{Materials available at:\href{https://github.com/kahramankostas/IoTDevIDv2}{github.com/kahramankostas/IoTDevIDv2}}.

The paper is organised as follows. Section~\ref{section:Related Work} reviews related work. Section~\ref{section:Materials and Methods} describes the methods we use to model network packets, and the datasets used for evaluation. Section~\ref{section:Model Selection} gives an overview of our model selection approach, Section~\ref{section:Performance Evaluation} evaluates the selected model, and Section~\ref{section:Performance Comparison} compares it against other DI approaches. Limitations are discussed in Section~\ref{section:Limitations}, and conclusions are presented in Section~\ref{section:Conclusion}.

\section{Related Work}\label{section:Related Work}

This section reviews previous studies that used fingerprinting to classify IoT devices. Fingerprints are feature sets derived from network packets or statistics to reflect the behavior patterns of devices. 
There are many studies in the literature on DI using fingerprints, but their applicability to IoT devices is controversial, as these often focus on the physical layer or application layer where IoT has wide protocol variety~\cite{bezawada2018behavioral}. Hence, we focus here on research that is based on network packet behavior, including relevant information from each of the data link, network, and transport layers. 
Table~\ref{table:comparisonRW} summarises the previous studies in comparison to our study with the IoTDevID method (see Table~\ref{tab:compare_all} for a comparison of results). 

One of the first studies to use network packet features in a fingerprint method to identify IoT devices was \textit{IoT Sentinel}~\cite{miettinen2017iot}. This study used a behavior-based fingerprint to identify vulnerable devices and isolate them from the network, using data from Aalto University IoT device captures. The fingerprints specific to each device were created based on 23 features extracted from each of the first 12 packets for each device, resulting in a fingerprint with 276 values. These 12 packets do not exactly represent flow; they are sequential packets from the same MAC address. The features contain information about source and destination, packet properties, and protocols used. However, since it performs packet merging from MAC addresses, it cannot identify non-IP devices (This situation, which we call the \textit{transfer problem}, is explored in-depth in Section~\ref{section:IoT Data Selection}). Also, the \textit{IP address count} feature used in this study is unlikely to generalize beyond the Aalto dataset, since it refers to the number of devices each device communicates with, which is environment-dependent. In this study, 17 of 27 device types were detected with an identification accuracy of above 95\%, and 10 with an accuracy of around 50\%.

Two other notable studies also used data from the Aalto dataset. In the first~\cite{hamad2019iot}, a fingerprint consisting of 67 statistical features was created using information extracted from the headers of Ethernet, IP, UDP, and TCP packets of 20-21 consecutive packets. Unlike the others, this study used network statistics in addition to features obtained from the packets. 
If a device used in one network is moved to a different network, the features extracted from the network packet will not change, but the network statistics will change. Since the network statistics are based not only on devices but also on their interrelationships, it is unlikely that they will generalize well. This study classified the devices with 90.3\% accuracy using various ML methods (see Table~\ref{table:comparisonRW}). The second study~\cite {aksoy2019automated} obtained 95\% accuracy using 33 features selected by a genetic algorithm (GA) from 212 features extracted from the headers of individual packets.
However, this study selectively presented its results with only part of the 
dataset (23 out of 27 device classes) and it used features that are unlikely to generalise (e.g., IP ID, TCP acknowledgment, TCP sequence, source and destination port numbers).

\textit{IoTSense}~\cite {bezawada2018behavioral} used selected features of IoT Sentinel based on their own design assessment, namely 17 protocol-based features which reflect device behavior. They also added three payload-related features, notably payload length and entropy. This feature list was applied to five packets for each device to produce a 100-member fingerprint, as an average of five packets were found to make up a session in this study. The packets are presumably merged according to common MAC address, though this is not explicitly stated in the paper. As a result of this study, per device recall of 93--100\% and an average accuracy of 99\%  were achieved. While some comparisons are made with the work of IoT Sentinel, the evaluation of IoTSense used a much smaller number of devices (i.e., 10 vs.\@ 31). In addition, the IoTSense experiment set began with 14 devices, though only 10 devices were used for the evaluation as four devices did not produce sufficient data for the analysis approach that was used.

Sivanathan et al.~\cite{sivanathan2018classifying} used the data obtained from a reasonably large variety of 28 IoT devices (such as cameras, lights, plugs, motion sensors, appliances and health-monitors) during six months to classify IoT devices. Eight features were used for the classification: flow volume, flow duration, average flow rate, device sleep time, server port numbers, domain name server (DNS) queries, network time protocol (NTP) queries and cipher suites. 
As a result, 28 IoT devices were classified with an accuracy of 99.88\%. However, four devices from the dataset were not used, and some elements of the feature set were too specific, thereby not focusing on device behavior, e.g., port numbers, DNS queries, and cipher suites.

\begin{table}[htbp]
	\centering
	\caption{Comparison of related work. AB: adaptive boosting, DT: decision tree, Dtab: decision table, GB: gradient boosting, kNN: k-nearest neighbours, LDA: latent Dirichlet allocation, NB: na\"ive Bayes, OneR: one rule, PART:  partial C4.5 DT,
		RF: random forest, SVM: support vector machine.}
	\setlength{\tabcolsep}{3pt}
	\begin{tabular}{@{}lllll@{}}
		\toprule
		Study &  Dataset  & \multicolumn{1}{l}{Devices} & Algorithm(s) & \multicolumn{1}{l}{Key comparators} \\
		\midrule
		\cite{miettinen2017iot}   &  Aalto  & 27    & RF    & \multicolumn{1}{l}{Transfer problem} \\
		\cite{hamad2019iot} &  Aalto  & 27    & RF, kNN, GB, DT & \multicolumn{1}{l}{Transfer problem,} \\
		&       &       & NB, SVM, AB, LDA & \multicolumn{1}{l}{overly specific features} \\
		\cite{aksoy2019automated}    & \multicolumn{1}{l}{Aalto } & 23    &DT, Dtab & \multicolumn{1}{l}{Used partial dataset,} \\
		&       &       & PART, OneR & \multicolumn{1}{l}{overly specific features} \\
		\cite{bezawada2018behavioral}     & Private  & 10    & GB, DT, kNN & \multicolumn{1}{l}{Used partial dataset} \\
		\cite{sivanathan2018classifying}     & UNSW   & 28    & NB, RF & \multicolumn{1}{l}{Overly specific features} \\\hline
		Our   & Aalto, & 27    & DT, GB, kNN, & \multicolumn{1}{l}{No trans.\@ prob.\@ nor overly}  \\
		Study  & UNSW  & 33    & NB, RF, SVM     & \multicolumn{1}{l}{spec.\@ features. Tested on}  \\				
		& & & & \multicolumn{1}{l}{second, indep.\@ dataset}  \\
		\bottomrule
	\end{tabular}%
	\label{table:comparisonRW}%
\end{table}%

\section{Materials and Methods}\label{section:Materials and Methods}

\subsection{System Model and Design Goals}\label{section:System Model and Design Goals}
IoTDevID analyzes the behavior of IoT devices in order to identify their brand and model. The resulting normal behavior profiles can then be used to secure devices, e.g., by finding devices with known vulnerabilities. More details about the threat model can be found in Section \ref{section:Limitations}. Our approach has similarities to the work reviewed in the previous section, in that we use ML methods to train device fingerprints from information stored in packet headers. However, we approach this in a more systematic way, using feature selection to identify fingerprints that generalise in a meaningful way from the training data, and with an emphasis on developing a model that can be deployed within real network environments.

\subsection{IoT Data Selection} \label{section:IoT Data Selection}
We used two public datasets to measure and compare the performance of our DI approach. Both of these contain real device data.  Since our approach is designed for benign networks, the datasets do not contain attack data. 
The first, the Aalto University IoT Devices Captures 
dataset~\cite{miettinen2017iot} (referred to as \textit{Aalto dataset}) has 31 devices and contains only the device installation data, but this installation process was repeated 20 times for each device to increase the amount of data~\cite{miettinen2017iot}. Whilst there are a total of 31 devices in this dataset, four of these devices are in pairs (see the supplementary material SM-Table~\ref{tab:macs}). For example, there are two WeMoSwitches with different MAC/IP addresses. Our purpose is to detect according to device behavior, so these device pairs are considered as a single device, not as two separate devices. Therefore, although the dataset contains 31 devices, it contains 27 classes.  Another characteristic of the dataset is that two devices using low energy protocols connect to the gateway where the data is collected via other devices. Therefore, these two devices (D-LinkDoorSensor and HueSwitch) do not have identifying features such as 
their own MAC and IP address. They use the IP/MAC address of the devices they communicate through (D-LinkHomeHub and HueBridge respectively). We refer to this as ``the  transfer problem'' in this paper (related device MAC addresses and labels are shared in SM-Table~\ref{tab:macs}). Notably, when using an individual packet-based approach (which is independent of MAC/IP addresses), our method is able to identify devices that are suffering from the transfer problem.

The second, UNSW IoT Traffic Traces~\cite{sivanathan2018classifying}
(referred to as \textit{UNSW dataset}) contains the day-to-day network logs of 28 IoT devices, each recorded over a period of 26 weeks. However, only a 60-day portion of this data is publicly available. This limited data does not include the four devices (Ring Door Bell, Hello Barbie, August Doorbell Camera, Belkin Camera) from  the work of Sivanathan et al.\@~\cite{sivanathan2018classifying}.
Therefore, we extracted data for these devices from the \href{https://iotanalytics.unsw.edu.au/attack-data}{benign data} of another study~\cite{hamza2019detecting} by the same institution. 
The dataset we use contains a total of 32 IoT devices and 7 non-IoT devices. 
We gathered non-IoT devices under one class, as the other study~\cite{sivanathan2018classifying} working with this dataset did, resulting in 33 total labels. Since the size of the UNSW dataset is quite large, we used the smaller Aalto dataset to formulate our modelling approach 
and the UNSW dataset to measure its broader generality.

\subsection{Individual, Aggregated and Mixed Method Packets}\label{section:Using Individual} 

Creating 
fingerprints by combining multiple packets is a common strategy~\cite{miettinen2017iot,hamad2019iot,bezawada2018behavioral,sivanathan2018classifying}, where identifying features such as a MAC or IP address are used to combine multiple packets.
These features uniquely identify the device, allowing us to know the source of the data,  but do not provide any information about the device type or behavior. Assuming that packets from the same address come from the same device, larger fingerprints are often constructed by combining these packets. However, this assumption is not always correct. In cases where there is a \emph{transfer problem} (e.g., multiple devices connecting through the same gateway), an IP/MAC address can represent more than one device. Consequently, the packet merging logic will not work, as devices with the same MAC address but different behavior will be considered as a single device. Further, there is no standard for the size of this merging process (number of packets, duration, etc.). Without merging,  the use of individual packets is beneficial in terms of speed and efficient use of limited resources~\cite{hwang2019lstm}. For these reasons, unlike many previous studies~\cite{miettinen2017iot,bezawada2018behavioral,sivanathan2018classifying,hamad2019iot}, we chose to use \emph{individual packets}, not merged ones as the basis for discrimination.

However, when viewed individually, some individual packets may be ambiguous and match the behavioral profile of more than one device. Therefore, the success rate from discrimination based on single packets is limited. We address this by introducing an alternative to individual packets, called \emph{aggregated method} that groups packets based on identifier attribute (MAC address) and ML labels assigned at the individual packet level. In this way, we can increase the success rate by ignoring the mislabelled packets. The difference of this approach from previous work which merged packets is that the behavior analysis in our aggregation  is performed \textit{before} the merging process. In other words, we combine not the packets, but the labels assigned to these packets by the ML algorithm. 
Since, this method does not work for MAC addresses suffering from the transfer problem, we have added an exception to our evaluation phase and followed a \emph{mixed method} to deal with this issue. This reevaluation process works in parallel with the pseudocode (see SM-Algorithm~\ref{alg:c_a} for pseudocode and corresponding line numbers mentioned below) as follows. 

Our aggregation algorithm  takes the device MAC addresses ${M}=\{{m}_{1}, {m}_{2},\ldots,{m}_{n}\}$, the result of the ML algorithm (initial labels) $\hat{Y}=\{\hat{y}_{1}, \hat{y}_{2},\ldots,\hat{y}_{n}\}$, and group size ($g$) as input.
The labels originating from the same MAC address are collected in the same array (lines 4-12).  Each array is divided into small pieces of group number size (lines 13-16). Aggregated results are created by assigning the mode (the most frequent element) to the entire group as a new label $\hat{Y}'=\{\hat{y}'_{1}, \hat{y}_{2}',\ldots,\hat{y}'_{n}\}$ (lines 17-23). On the other hand, for creating the exception list of aggregation process, the MAC address corresponding to each label is added to the same array (lines 33-36).  If a MAC address is the most frequent element in more than one array, it is added to the exception list (lines 37-41). As mixed result, individual labels ($\hat{y}$) are applied for the MAC addresses in the exception list, and aggregated labels ($\hat{y}'$)  are applied for the remainder  (lines 42-47).  

However, while the aggregation algorithm works well for DI with benign data, care should be taken in networks with malicious data. Depending on the tolerance of the aggregation algorithm, malicious packets that impersonate an IP/MAC address and are close to the behavior of the benign packets with the same IP/MAC address may be grouped together (see Limitations~\ref{section:Limitations}).

\subsection{Feature Extraction }
\label{sec:featureSet}
We performed feature extraction from  the packet headers of \texttt{pcap} (packet capture) files. Before this, we separated the \texttt{pcap} files into test and training sets to completely isolate 
these sets.
The Aalto dataset has 20 sessions (\texttt{pcap} files) 
per
device. We divided them in two parts: 16 parts training and four parts test data (80\%:20\%). The UNSW dataset consists of a different \texttt{pcap} file for each day's records. 
To isolate,
we constructed the training and test data from data collected on different days (\texttt{pcap} files) (see SM-Table~\ref{tab:unsw_features} for distribution). 
Section~\ref{section:Feature Selection} explains our feature selection process in detail.

In addition to the 111 features extracted from the network packet headers, we included payload entropy~\cite{bezawada2018behavioral}, protocol (from TCP-IP layers), source 
and destination port class~\cite{miettinen2017iot}, which are all features found to be useful in previous studies. While payload entropy gives clues about the characteristic of the payload, the protocol and port class features provide summary information about source and destination ports. Source-destination port class features gather port numbers under 14 classes: No port, 0-Reserved, 53-DNS, 67-BOOTP server, 68-BOOTP client, 80-HTTP, 123-NTP, 443-HTTPS, 1900-SSDP, 5353-mDNS,
49153-ANTLR, 0:1023-well-known ports, 1023:49151-registered ports, and 49151:65535-dynamic (private) ports~\cite{miettinen2017iot}. 
The port classification above is ordered so that, for example, a packet with port 53 is not also classified as a well-known port. 
Section~\ref{section:Feature Selection} explains further how port number classes are created. 

MAC/IP addresses are source-destination based identifying features. Although they uniquely identify source and destination devices, they do not provide information about device behaviors. For example, two devices with the same behavior can have different MAC and IP addresses. Because 
they
are not predictable and generalizable, we do not use them in our feature set (though they are used in our aggregation method). 

\section{Model Selection}\label{section:Model Selection}
\subsection{Feature Selection}\label{section:Feature Selection}
\begin{figure}[ht]
	\centerline{\includegraphics[width=1\columnwidth]{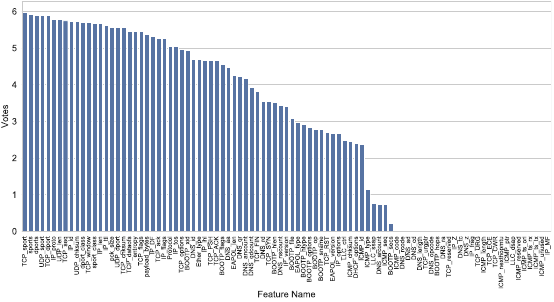}}
	\caption{Results of feature importance voting on the Aalto dataset. Features rated with six different feature importance methods are listed from the most-to-least voted. The highest ranked features are related to size, payload characteristics and frequently used protocols such as IP, UDP, TCP,  while the lowest ranked features are related to rarely used features of protocols.}
	\label{fig:voting}
\end{figure}

With our initial features extracted, we used a feature-importance-based voting method via the xverse\ (\href{https://pypi.org/project/xverse/}{pypi.org/project/xverse/}) package to eliminate unnecessary features. This method calculates the importance scores of all features for each device using six different techniques, and then uses voting to decide whether to include them (see Fig.~\ref{fig:voting} for features and their vote rates). The six scoring techniques used by this method are information value using the weight of evidence, variable importance using RF, recursive feature elimination, variable importance using extra trees classifier, chi-square best variables, and L1-based feature selection. We removed from our feature pool the 26 features that failed to receive votes on any device from any of these six techniques.  

A number of the remaining features contain information that is potentially specific to a particular session, and therefore 
may not be
useful for identifying devices more generally. For some features, this is clear, e.g., the initial values of the IP ID, TCP sequence and acknowledgment numbers are randomly assigned, and the next values are consecutive numbers following this initial value; hence, they they do not contain information that is useful for DI beyond their current session. For other features, such as TCP port numbers, their generality is less clear. So, to measure whether or not each of these features is beneficial, we excluded all of them from the feature set and trained an ML model (DT) to determine a baseline level of performance. In turn, we then added each of the features to the baseline set and observed whether this improved or impaired the performance.

\begin{figure}[ht]

	\centerline{\includegraphics[width=1\columnwidth]{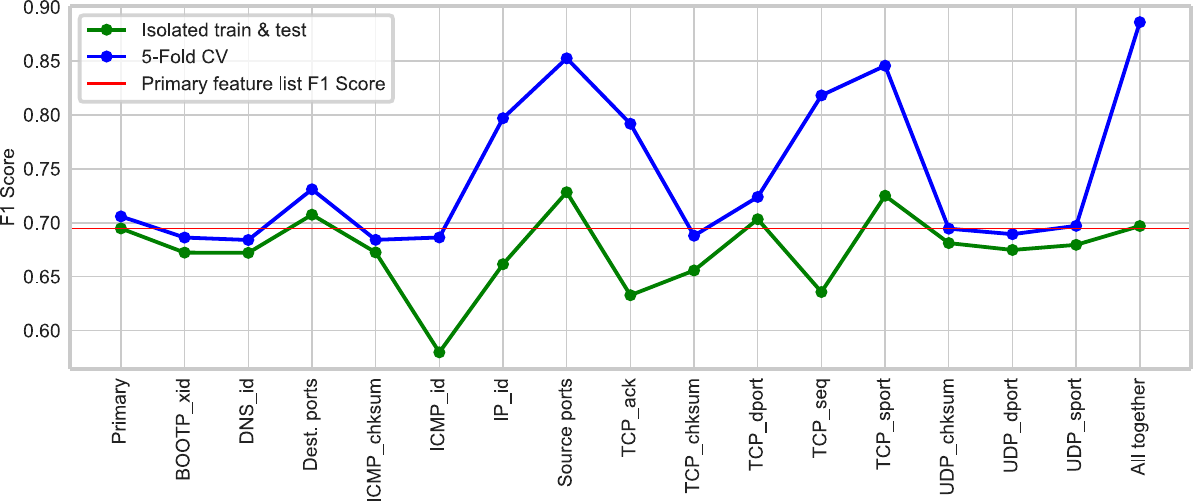}}
	\caption{Generality of identifying features, measured using  isolated test sets and cross-validation. It is seen that, due to data leakage from  training set to test set, cross-validation considerably over-estimates the value of these features.}
	\label{fig:comparedata}
\end{figure}

The results are shown in Fig.~\ref{fig:comparedata}. First of all, it is notable that the results are strongly dependent on whether the models are evaluated using cross-validation or using isolated training and test sets. This is presumably because, when using cross-validation, it is more likely that the training and test data will be sampled from the same session --- meaning that session-specific patterns learned in the training set will also generalise to the test set. This may explain why session-specific features were chosen during feature selection, and highlights the danger of using off-the-shelf feature selection algorithms on data of this kind. Using isolated training and test sets, by comparison, it is clear that session-specific features such as the IP ID cause over-fitting and impair the generality of models.

\begin{figure}[ht]

	\centerline{\includegraphics[width=0.9\columnwidth]{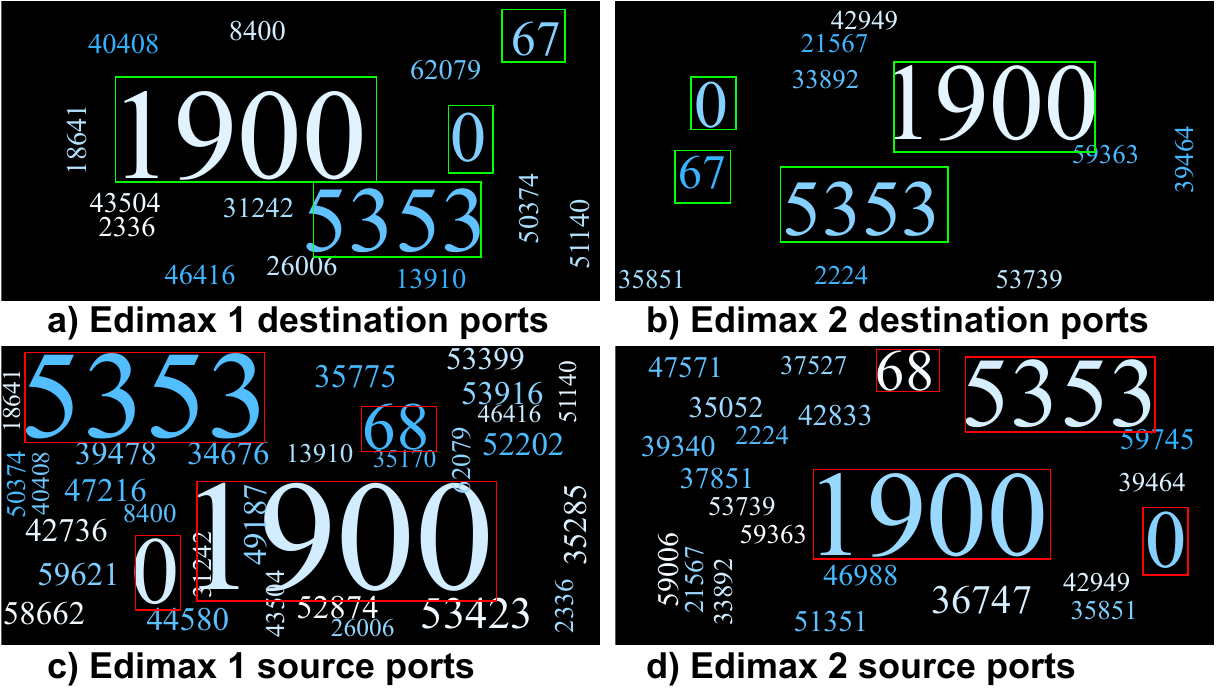}}
	\caption{Word clouds of port numbers used by two devices of the same type, with digit size showing frequency of use. Red and green boxes, showing port numbers specific to protocols such as SSDP, mDNS, and BOOTP, generalise between device instances. The other port numbers do not generalise.}
	\label{fig:ports}
\end{figure}

On the other hand, Fig.~\ref{fig:comparedata} shows that port-based features can be useful for building generalisable models. 
For example, consider the port numbers used by one of the devices in the Aalto dataset which contains multiple device instances. 
Fig.~\ref{fig:ports} shows a word cloud built from the data from the two Edimax camera instances. 
Although a few port numbers (representing the protocols BOOTP, HTTP, SSDP and mDNS) generalise between the devices, the remaining ports are presumably session-based and do not generalise. To address this, we do not use the raw feature values for port-based features, but rather map them to discrete values representing port classes and protocols (see Section~\ref{sec:featureSet}) --- in effect, pruning the instance-specific information whilst preserving the behavioral information they provide.

Finally, after eliminating identifying and redundant features, we used a GA to decide on the most appropriate feature set from the remaining 52-member feature pool. The GA uses a wrapper method and thus tests the usefulness of feature sets in the context of a particular classifier (DT). In this way, we have created a feature subset that detects with higher performance, while decreasing the model complexity by reducing the number of features. Table~\ref{tab:feature_list} shows (highlighted in green) the final list of features selected by the GA.

\begin{table}
	\centerline{\includegraphics[width=1\columnwidth]{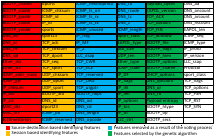}}
	\caption{Results of feature selection. Red and orange features were removed due to a lack of generality. Blue features were removed as a result of feature importance voting. Of the remaining features, those in green were identified by the GA as being most useful for discriminating between devices. They include size, payload and fields related to common protocols.}
	\label{tab:feature_list}
\end{table}

\subsection{Algorithm Selection}\label{section:ML Algorithm Selection}
Based on earlier approaches (see Section~\ref{section:Related Work}) and surveys of related work~\cite{hussain2020machine, al2020survey}
which sought to train an ML algorithm to correctly predict a device's type from extracted features,
we considered the following six ML algorithms: RF, kNN, GB, DT, NB, and SVM. We used
random search with nested cross-validation (as implemented by scikit-learn:~\href{https://scikit-learn.org/stable/}{scikit-learn.org/stable/}) to find suitable hyperparameters for each algorithm. Using these algorithms, we trained multi-class classifiers to discriminate the 27 classes from the Aalto dataset, training each model 100 times to measure its stability. See Table~\ref{table:ml-result1} for the results.

\begin{table}[ht]
	\scriptsize	
	
	\caption{Comparison of ML algorithms with average and standard deviation (SD) of 100 repeats on the Aalto dataset. t is time, in seconds.}
	\centering
	\setlength{\tabcolsep}{3pt}
	\begin{tabular}{@{}lllllll@{}}
		
		\toprule
		ML    & Accuracy & Precision &  Recall  & F1 score & Train-t & Test-t \\
		\midrule
       DT    	& 0.705±0.001 	& 0.774±0.003 	& 0.706±0.001 	& \underline{0.727}±0.001 	& 0.128 	& \underline{0.004} \\
    GB    	& 0.699±0.001 	& \underline{0.789}±0.003 	& 0.693±0.001 	& 0.725±0.001 	& 918.3 	& 8.312 \\
    kNN   	& 0.705±0.000 	& 0.752±0.000 	& 0.705±0.000 	& 0.718±0.000 	& \underline{0.005} 	& 20.20 \\
    NB    	& 0.617±0.000 	& 0.584±0.000 	& 0.629±0.000 	& 0.559±0.000 	& 0.433 	& 0.032 \\
    RF    	& \underline{0.708}±0.001 	& 0.768±0.004 	& 0.708±0.001 	&\underline{0.727}±0.002 	& 3.742 	& 0.333 \\
    SVM   	& 0.680±0.000 	& 0.697±0.000 	& 0.634±0.000 	& 0.649±0.000 	& 101.3 	& 64.80 \\ 

		\bottomrule
	\end{tabular}
	\label{table:ml-result1}  
	
\end{table}												

It can be seen that RF, DT and GB are the best performing algorithms, in terms of DI. The fastest in terms of inference time are DT and NB, though the accuracy of NB is very low. Although kNN and GB perform relatively well in terms of accuracy, they are not practical to use due to their slowness. The SVM algorithm is also not a reasonable option in terms of speed and accuracy. kNN, GB, and SVM are particularly disadvantageous due to their very large inference times. In a real-time device detection system that will operate in a millisecond-level processing environment such as network traffic, inference time in seconds are not acceptable. Based on these observations, we use DT in the remainder of this work, since it offers the best balance between speed and accuracy.

\section{Performance Evaluation}\label{section:Performance Evaluation}

\begin{figure}[ht]
	\centerline{\includegraphics[width=.9\columnwidth]{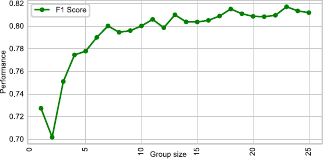}}
	\caption{Relationship between aggregation group size and performance. The benefit of increasing group size appears to plateau around $g=13$.}
	\label{fig:size}
\end{figure}

Based on our model selection we present our results here for the three variations of our method: individual, aggregated, and mixed. First, we looked at the relationship between group size and performance within the context of the aggregation algorithm. From Fig.~\ref{fig:size}, there is a positive relationship, showing that larger group sizes are generally more effective. However, many of the IoT devices communicate infrequently, so large group sizes would be impractical in some cases. For this reason, we selected a group size of 13, which is approximately the point where performance starts to plateau.

Table~\ref{tab:results} shows overall results for the two datasets. Using the individual packet approach, an F1 score of 73\% was achieved in the Aalto dataset and 83\% in the UNSW dataset. However, a significant increase in the ability to correctly identify devices of both datasets was observed by using aggregation. The overall F1 score increased to 81\% on the Aalto dataset and to approximately 94\% on the UNSW dataset. Since feature selection was done using only the Aalto dataset, it is very encouraging to see such a high level of discrimination on the UNSW dataset. This is a good indication that the selected feature set will generalise well to other IoT environments.

\begin{table}[htbp]
  \centering
  \caption{Results and their SD obtained using Individual, Aggregated and Mixed approaches on Aalto and UNSW datasets. t is time in seconds. Alg-t indicates the running time of the aggregation algorithm.}
  	\setlength{\tabcolsep}{3pt}
    \begin{tabular}{@{}llcccc@{}}
    \toprule
    Method & Dataset & Accuracy &  F1 score & Test-t & Alg-t \\
    \midrule
    \multirow{2}[1]{*}{Individual} & Aalto & 0.705±0.001 & 0.727±0.001 & 0.004 & 0.000 \\
          & UNSW  & 0.853±0.010 & 0.834±0.012 & 0.008 & 0.000 \\
\cmidrule{2-6}     \multirow{2}[0]{*}{Aggregated} & Aalto & 0.745±0.011 & 0.809±0.005 & 0.007 & 0.164 \\
          & UNSW  & 0.943±0.012 & 0.937±0.017 & 0.017 & 0.425 \\
 \cmidrule{2-6}   \multirow{2}[1]{*}{Mixed} & Aalto & 0.833±0.002 & 0.861±0.004 & 0.008 & 0.216 \\
          & UNSW  & 0.941±0.012 & 0.935±0.017 & 0.022 & 0.479 \\
    \bottomrule
    \end{tabular}%
  \label{tab:results}%
\end{table}%

Table~\ref{tab:device} shows the average discrimination performance of DT models at the device level for the Aalto Dataset. As Table~\ref{tab:device} also shows, the dataset is highly imbalanced, meaning that we need to use a metric that is relatively insensitive to class size imbalance in order to give a meaningful picture of the model's performance at the device level -- hence our use of the F1 score. Other works have used accuracy for this, which is an inappropriate metric for imbalanced datasets.

\begin{table}[tb]
  \centering
  \setlength{\tabcolsep}{3pt}
  \caption{Device proportions over the entire dataset, F1 scores per device according to different approaches (Aalto Dataset, average of 100 repetitions).}
    \begin{tabular}{@{}lrcccc@{}}
    \toprule
    \multirow{2}[4]{*}{Device name} & \multicolumn{2}{c}{Packet statistics} & \multicolumn{3}{c}{Packet discrimination} \\\cline{2-3}\cline{4-6}         & \multicolumn{1}{c}{Packets} & Percent & Individual & Aggregated & Mixed \\
    \midrule
    Aria  	& 441   	& 0.420 	&	0.932&	1.000&	1.000	\\
    D-LinkCam 	& 6244  	& 5.940 	&	0.891&	1.000&	0.988	\\
    D-LinkDayCam 	& 1063  	& 1.010 	&	0.864&	1.000&	1.000	\\
    D-LinkDoorSensor 	& 1892  	& 1.800 	&	0.762&	0.057&	0.788	\\
    D-LinkHomeHub 	& 8595  	& 8.180 	&	0.681&	0.797&	0.777	\\
    D-LinkSensor 	& 6549  	& 6.230 	&	0.382&	0.644&	0.626	\\
    D-LinkSiren 	& 6186  	& 5.890 	&	0.367&	0.640&	0.631	\\
    D-LinkSwitch 	& 6519  	& 6.200 	&	0.665&	0.969&	0.964	\\
    D-LinkWaterSensor 	& 6435  	& 6.120 	&	0.392&	0.624&	0.622	\\
    EdimaxCam 	& 831   	& 0.790 	&	0.872&	1.000&	1.000	\\
    EdimaxPlug1101W 	& 1160  	& 1.100 	&	0.601&	0.827&	0.824	\\
    EdimaxPlug2101W 	& 1010  	& 0.960 	&	0.453&	0.708&	0.699	\\
    EdnetCam 	& 408   	& 0.390 	&	0.833&	1.000&	1.000	\\
    EdnetGateway 	& 683   	& 0.650 	&	0.908&	1.000&	1.000	\\
    HomeMaticPlug 	& 611   	& 0.580 	&	1.000&	1.000&	1.000	\\
    HueBridge 	& 13936 	& 13.260 	&	0.810&	0.217&	0.815	\\
    HueSwitch 	& 18448 	& 17.560 	&	0.891&	0.720&	0.891	\\
    IKettle2 	& 145   	& 0.140 	&	0.727&	1.000&	1.000	\\
    Lightify 	& 4149  	& 3.950 	&	0.977&	1.000&	1.000	\\
    MAXGateway 	& 567   	& 0.540 	&	0.964&	1.000&	1.000	\\
    SmarterCoffee 	& 149   	& 0.140 	&	0.727&	0.983&	0.981	\\
    TP-LinkPlugHS100 	& 667   	& 0.630 	&	0.693&	0.748&	0.746	\\
    TP-LinkPlugHS110 	& 636   	& 0.610 	&	0.429&	0.336&	0.328	\\
    WeMoInsightSwitch 	& 5962  	& 5.670 	&	0.667&	0.874&	0.866	\\
    WeMoLink 	& 6625  	& 6.300 	&	0.638&	0.929&	0.924	\\
    WeMoSwitch 	& 4477  	& 4.260 	&	0.511&	0.769&	0.768	\\
    Withings 	& 688   	& 0.650 	&	1.000&	1.000&	1.000	\\

    \bottomrule
    \end{tabular}%
  \label{tab:device}%
\end{table}%

\begin{table}[tb]
  \centering
  \setlength{\tabcolsep}{3pt}
  \caption{Device proportions over the entire dataset, F1 scores per device according to different approaches (UNSW Dataset, average 100 repetitions).}
    \begin{tabular}{@{}lrcccc@{}}
    \toprule
    \multirow{2}[4]{*}{Device name} & \multicolumn{2}{c}{Packet statistics} & \multicolumn{3}{c}{Packet discrimination} \\\cline{2-3}\cline{4-6}         & \multicolumn{1}{c}{Packets} & Percent & Individual & Aggregated & Mixed \\
    \midrule
 Amazon Echo 	&	10000	&	2.831	&	0.882	&	0.985	&	 0.983 \\
AugustDoorbellCam 	&	10000	&	2.831	&	0.701	&	0.960	&	 0.955 \\
AwairAirQualityMon	&	10000	&	2.831	&	0.952	&	0.993	&	 0.994 \\
Belkin Camera 	&	10000	&	2.831	&	0.890	&	1.000	&	 1.000 \\
BelkinWeMoSwitch 	&	10000	&	2.831	&	0.803	&	0.988	&	 0.988 \\
BelkinWeMoSensor 	&	10000	&	2.831	&	0.772	&	0.987	&	 0.988 \\
BlipcareBPMeter 	&	119	&	0.034	&	0.851	&	1.000	&	 1.000 \\
Canary Camera 	&	10000	&	2.831	&	0.761	&	0.928	&	 0.911 \\
Dropcam 	&	10000	&	2.831	&	0.998	&	1.000	&	 1.000 \\
GoogleChromecast 	&	10000	&	2.831	&	0.448	&	0.347	&	 0.341 \\
HPPrinter 	&	10000	&	2.831	&	0.468	&	0.710	&	 0.707 \\
HelloBarbie 	&	164	&	0.046	&	0.677	&	1.000	&	 1.000 \\
InsteonCam	&	10000	&	2.831	&	0.953	&	0.997	&	 0.997 \\
LightLiFXSmartBulb 	&	10000	&	2.831	&	0.979	&	1.000	&	 1.000 \\
NEST-PSmokeAlarm 	&	7063	&	1.999	&	0.935	&	1.000	&	 1.000 \\
Nest Dropcam 	&	10000	&	2.831	&	0.966	&	1.000	&	 1.000 \\
NetatmoWelcome 	&	10000	&	2.831	&	0.922	&	0.998	&	 0.994 \\
NetatmoWeatherSt	&	10000	&	2.831	&	0.896	&	1.000	&	 1.000 \\
Non-IoT 	&	67295	&	19.05	&	0.872	&	0.929	&	 0.929 \\
PIX-STAR-Pframe 	&	10000	&	2.831	&	0.587	&	0.928	&	 0.885 \\
PhillipHLightbulb 	&	10000	&	2.831	&	0.971	&	1.000	&	 1.000 \\
RingDoorBell 	&	10000	&	2.831	&	0.856	&	0.999	&	 0.995 \\
SamsungSCam 	&	10000	&	2.831	&	0.678	&	0.892	&	 0.896 \\
SmartThings 	&	10000	&	2.831	&	0.995	&	1.000	&	 1.000 \\
TP-LinkCloudCam	&	10000	&	2.831	&	0.979	&	1.000	&	 1.000 \\
TP-Link Smart plug 	&	10000	&	2.831	&	0.875	&	1.000	&	 1.000 \\
TP-LinkRouter	&	10000	&	2.831	&	0.908	&	0.971	&	 0.979 \\
TribySpeaker 	&	10000	&	2.831	&	0.908	&	0.999	&	 0.999 \\
WithingsSlpSensor 	&	10000	&	2.831	&	0.437	&	0.404	&	 0.411 \\
WithingsBabyMon 	&	10000	&	2.831	&	0.984	&	1.000	&	 1.000 \\
WithingsSScale 	&	8279	&	2.344	&	0.973	&	1.000	&	 1.000 \\
iHome 	&	10000	&	2.831	&	0.958	&	1.000	&	 1.000 \\
unknownIoT 	&	340	&	0.096	&	0.676	&	0.894	&	 0.891 \\

    \bottomrule
    \end{tabular}%
  \label{tab:deviceUNSW}%
\end{table}%

When the device-based results of the Aalto dataset are examined, we see that the aggregation algorithm contributes positively to almost all devices, while negatively affecting only four devices. This is because the pairs that make up these four devices suffer from the transfer problem (they share the same MAC address, see SM-Table~\ref{tab:macs}). To deal with this, we used our mixed method by adding an exception to our aggregation algorithm. 
Table~\ref{tab:results} shows that with the application of the mixed approach, the overall F1 score, which was 81\% in the Aalto dataset, increased to 86\%. Since there is no transfer problem in the UNSW dataset, this approach does not impact its results significantly --- see Table~\ref{tab:deviceUNSW}.

\begin{figure}[ht]
	\centerline{\includegraphics[width=0.9\columnwidth]{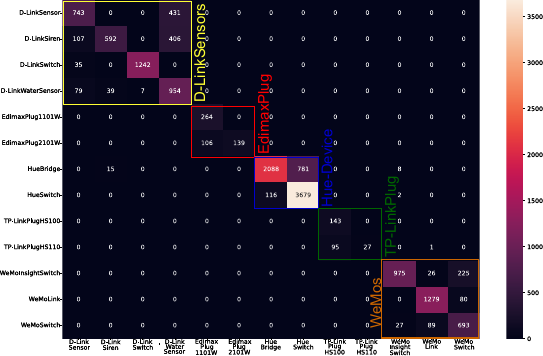}}
	\caption{Confusion matrix of devices with poor discrimination. The coloured groups show this is mostly due to the difficulty of discriminating between similar devices from the same manufacturer.}
	\label{fig:cm_group}
\end{figure}

It is notable that the performance on the Aalto dataset is significantly lower than the UNSW dataset. This appears to be due to certain device subgroups. A confusion matrix containing only low-performance devices is given in Fig.~\ref{fig:cm_group}.
In this case, the devices in a subgroup have some similarities:
they are either similar purpose devices manufactured by the same companies (e.g. yellow, blue and orange groups in Fig.~\ref{fig:cm_group}) or different models of the same device (e.g. red and green groups in Fig.~\ref{fig:cm_group}). It does not seem possible to perfectly separate these devices according to their behavior, at least when observed at the network level. However, it is likely that these devices use very similar hardware and software, and so exhibit similar behavior, 
as well as similarities in vulnerabilities and their prevention~\cite{miettinen2017iot}.  
Consequently, from a DI perspective, it is plausible to consider these devices under a single label. 
When doing this, 
the accuracy for the Aalto dataset increases from 73\% to 88\% for the individual method, and  from 86\% to 97\%  for the mixed method.

\section{Comparison with previous work}\label{section:Performance Comparison}

In Table~\ref{tab:compare_all}, we compare the overall results of our study with previously published results. While, at first glance, our study does not appear to yield higher numeric results than existing work, there are a number of reasons why this is not a fair comparison. First, many approaches use overly specific features that are unlikely to generalize to unseen datasets. As our analysis in Section~\ref{section:Feature Selection} showed, this may be a result of information leaking from the test set into the training set, something that is easy to do unintentionally. Second, there is widespread use of accuracy figures with imbalanced datasets; since IoT device datasets inherently suffer from unbalanced distributions, using accuracy as the sole evaluation criterion can be highly misleading. Third, there is inconsistency amongst the datasets used, with some studies reporting results for only partial datasets. We summarize these points in Section~\ref{section:Related Work} and Table~\ref{table:comparisonRW}. In our work, by comparison, we were careful to filter out dataset-specific features, to prevent information leakage, to use appropriate metrics, and to report full results. We also explicitly used a second dataset to measure the generality of the approach. Consequently, we believe that our results are unbiased and more likely to indicate likely success on unseen data than those reported in previous studies.

\begin{table}[htbp]
	\centering
	\caption{Comparing the result of IoTDevID with former studies. Table~\ref{table:comparisonRW} summarises issues with results of previous studies.}
	\setlength{\tabcolsep}{3pt}
	\begin{tabular}{@{}lllll@{}}
		\toprule
\multicolumn{1}{l}{Study} & Dataset & \multicolumn{1}{l}{Result} & \multicolumn{1}{l}{Metric}  & \multicolumn{1}{l}{Feature Types} \\
\midrule
\multicolumn{1}{l}{\cite{miettinen2017iot}} & Aalto & 81.50\% & \multicolumn{1}{l}{Accuracy} & \multicolumn{1}{l}{Packet header  } \\
\multicolumn{1}{l}{\cite{hamad2019iot}} & Aalto & 90.30\% & \multicolumn{1}{l}{F1 score}  & \multicolumn{1}{l}{Flow statistics  } \\

\multicolumn{1}{l}{\cite{aksoy2019automated}} & Aalto & 82\%  & \multicolumn{1}{l}{Accuracy}  & \multicolumn{1}{l}{Packet header  } \\

\multicolumn{1}{l}{\cite{bezawada2018behavioral}} & Private & 99\%  & \multicolumn{1}{l}{Accuracy} & \multicolumn{1}{l}{Packet header \& Payload} \\

\multicolumn{1}{l}{\cite{sivanathan2018classifying}} & UNSW  & 99.98\% & \multicolumn{1}{l}{Accuracy}  &\multicolumn{1}{l}{Packet header \& Flow} \\
\midrule
& Aalto & 83.30\% & \multicolumn{1}{l}{Accuracy} &  \multicolumn{1}{l}{Packet header \& Payload }        \\
\multicolumn{1}{l}{Our} & UNSW  & 94.30\% &       & \\
\cmidrule{2-5} \multicolumn{1}{l}{Study} & Aalto & 86.10\% & \multicolumn{1}{l}{F1 score}  & \multicolumn{1}{l}{Packet header \& Payload } \\
& UNSW  & 93.70\% &       &        \\
\bottomrule
	\end{tabular}%
	\label{tab:compare_all}%
\end{table}%

A key contribution of our work is the determination of a feature set that provides good discrimination across datasets, and is therefore likely to provide a good basis for DI more generally. To give more insight into this, and to offer a more meaningful comparison against previous approaches, we compared this feature set against the feature sets used in IoTSense~\cite{bezawada2018behavioral} and IoT Sentinel~\cite{miettinen2017iot}. Note we did not include the other methods listed in Table~\ref{tab:compare_all} in this comparison, either because they were flow-based~\cite{hamad2019iot,sivanathan2018classifying} or their feature set was not shared~\cite{bezawada2018behavioral}. We applied all three of our approaches (individual, aggregated, and mixed) using all three feature sets and both datasets. Table~\ref{tab:compare} shows that, in all cases, our feature set resulted in significantly better device discrimination than the previously published feature sets. It is also notable that the performance metrics we observed when using the feature set from IoTSense are significantly worse than that method's published figure of 99\% accuracy, which again highlights the problem of basing comparisons on published figures alone.

\begin{table}[htbp]
	\centering
	\caption{Comparison of feature sets on the Aalto and UNSW datasets.}
	\setlength{\tabcolsep}{3pt}
   \begin{tabular}{@{}lllllrr@{}}

      \toprule
       \multicolumn{2}{c}{Data}&   Method  & Accuracy &  F1 score & Test-t & Alg-t \\
    \midrule
      
     \multirow{6}[4]{*}{\begin{sideways}Individual\end{sideways}} & \multirow{3}[2]{*}{\begin{sideways}Aalto\end{sideways}} &  IoTDevID   &  \underline{0.705}±0.001  &  \underline{0.727}±0.071  & 0.004 &  N/A \\
          &       &  IoTSense   &  0.639±0.000  &  0.561±0.001  & 0.006 &  N/A \\
          &       &  IoTSentinel  &  0.700±0.001  &  0.602±0.002  & 0.008 & N/A \\
\cmidrule{2-7}          & \multirow{3}[2]{*}{\begin{sideways}UNSW\end{sideways}} &  IoTDevID   & \underline{0.853}±0.010 & \underline{0.834}±0.012 & 0.022 & N/A \\
          &       &  IoTSense   & 0.710±0.010 & 0.697±0.012 & 0.023 & N/A \\
          &       &  IoTSentinel  & 0.526±0.010 & 0.510±0.010 & 0.027 & N/A \\
    \midrule
    \multirow{6}[4]{*}{\begin{sideways}Aggregated\end{sideways}} & \multirow{3}[2]{*}{\begin{sideways}Aalto\end{sideways}} &  IoTDevID   &  \underline{0.745}±0.011  &  \underline{0.809}±0.005  & 0.007 & 0.164 \\
          &       &  IoTSense   &  0.671±0.003  &  0.657±0.004  & 0.007 & 0.142 \\
          &       &  IoTSentinel  &  0.671±0.003  &  0.639±0.005  & 0.007 & 0.136 \\
\cmidrule{2-7}          & \multirow{3}[2]{*}{\begin{sideways}UNSW\end{sideways}} &  IoTDevID   & \underline{0.943}±0.012 & \underline{0.937}±0.017 & 0.023 & 0.425 \\
          &       &  IoTSense   & 0.840±0.011 & 0.834±0.014 & 0.024 & 0.450 \\
          &       &  IoTSentinel  & 0.710±0.015 & 0.689±0.016 & 0.026 & 0.420 \\
    \midrule
    \multirow{6}[4]{*}{\begin{sideways}Mixed\end{sideways}} & \multirow{3}[2]{*}{\begin{sideways}Aalto\end{sideways}} &  IoTDevID   &  \underline{0.833}±0.002  &  \underline{0.861}±0.004  & 0.008 & 0.216 \\
          &       &  IoTSense   &  0.748±0.004  &  0.701±0.005  & 0.007 & 0.174 \\
          &       &  IoTSentinel  &  0.778±0.010  &  0.691±0.008  & 0.006 & 0.165 \\
\cmidrule{2-7}          & \multirow{3}[2]{*}{\begin{sideways}UNSW\end{sideways}} &  IoTDevID   & \underline{0.941}±0.012 & \underline{0.935}±0.017 & 0.022 & 0.479 \\
          &       &  IoTSense   & 0.844±0.010 & 0.835±0.012 & 0.024 & 0.487 \\
          &       &  IoTSentinel  & 0.700±0.017 & 0.679±0.018 & 0.027 & 0.528 \\
    \bottomrule
    \end{tabular}%
	\label{tab:compare}%
\end{table}%

\section{Limitations}
\label{section:Limitations}

In our study, we used all the IoT DI datasets that were publicly available at the time 
(see Section~\ref{section:IoT Data Selection}). However, IoT technologies are a rapidly developing field and new data is likely to have been released even as we continued our work. We encourage other researchers to use other datasets to further test the generalizability of our results and we provide scripts to make this possible. We used a broad selection of commonly-used ML algorithms (see Section~\ref{section:ML Algorithm Selection}), though this was not exhaustive, and there may be other algorithms that might perform better. However, we have done some preliminary research using deep neural networks (not reported here), and so far our results suggest that more traditional ML techniques work better on this particular problem.

Although IoTDevID identifies devices well, it does not provide solutions for detecting vulnerable devices and taking necessary precautions about these devices. A Software-Defined Networking (SDN) based network management solution may be useful for this process. Furthermore, IoTDevID is currently intended only for use in benign networks, meaning that there is still a need for an intrusion detection system to protect against attacks. A further consideration is how a DI model would be deployed, trained and maintained in a particular network. Whilst we focus on the performance of multi-class classifiers, i.e., a single model for discriminating between all devices on the network, this may not be optimal in situations where devices are frequently added and removed. A more practical architecture in this case may be an ensemble of single-class classifiers, since this would not require retraining of the entire model. However, our initial results (not reported here) suggest there may be a trade-off against performance, with multi-class models offering better discrimination.

\section{Conclusions}\label{section:Conclusion}

In this study, we present an ML-based method for identifying IoT devices. Identifying the IoT devices with a network is important, both for finding and removing rogue devices, and for understanding the security vulnerabilities of a network more generally. A number of previous studies have attempted to use ML in this process. However, we have identified various factors that may limit the generality of their findings, including the use of session-based identifying features, and the use of inappropriate metrics. In this study, we have attempted to go about this process in a rigorous and transparent manner, with the aim of developing a robust method of DI that reliably generalises beyond the data on which it was trained.

A key contribution of this work is the use of multi-stage feature selection to determine a set of generalizable packet-level features that can be used for building robust models. Notably, we compared this feature set against those used in previous studies, and showed that it supports the training of significantly better ML models. We also demonstrate that the feature set generalises well to a dataset that was unseen during feature selection, giving us confidence that it provides a meaningful basis for the broader discrimination of IoT devices.

We also address the problem of detecting non-IP devices. A limitation of earlier approaches is that they use IP or MAC addresses to merge packets prior to applying an ML model, which means they are not applicable in situations where a device does not have an IP or MAC address. Instead, we apply ML at the packet level, and then use an aggregation algorithm that takes into account both the packet-level classification and, if available, the IP or MAC address. If the latter is not available, it is still possible to carry out identification based upon individual packets alone.

From an ML-perspective, we found decision trees to offer the best trade-off between predictive performance and inference time, the latter being an important consideration for deploying a model to monitor network traffic in real time. 

In future work, we plan to develop an IDS that will work with IoTDevID and detect attacks against the network, and an SDN-based network management system that evaluates both device identification and intrusion detection outputs. Thus, the aim is to have an IoT security system that avoids risk of potential vulnerabilities, prevents attacks and can be used in a real world setting.

\bibliographystyle{IEEEtran}
\bibliography{references}
\begin{IEEEbiography}[{\includegraphics[width=.8in,height=1.2in,clip,keepaspectratio]{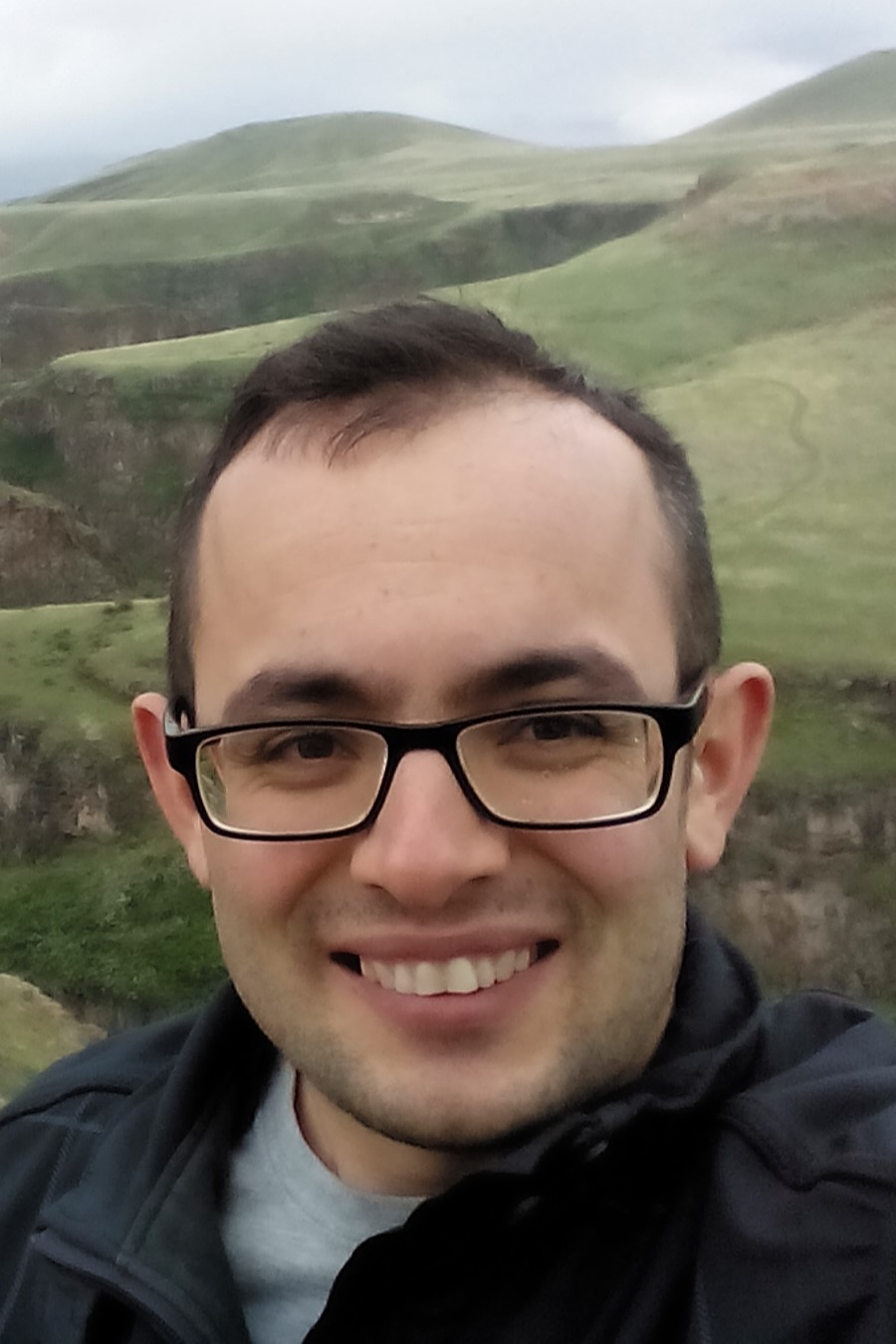}}]{Kahraman Kostas} 
received  the MSc degree  in  Computer Networks and Security from  the  University of Essex, Colchester, U.K., in 2018. He is a PhD candidate in Computer  Science at  Heriot-Watt University, Edinburgh, U.K. His research focuses on the security of computer networks and Internet of Things.  You can find more information at \url{https://kahramankostas.github.io/}.\end{IEEEbiography}

\begin{IEEEbiography}[{\includegraphics[width=0.8in,height=1.2in,clip,keepaspectratio]{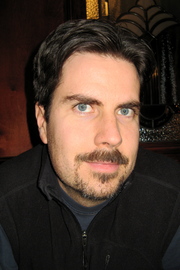}}]{Mike Just} earned his Ph.D. in Computer Science from Carleton University in 1998 and is currently an Associate Professor at Heriot-Watt University. He is primarily interested in computer security, and in applying human-computer interaction and machine learning techniques to solve computer security problems. You can find more information at \url{https://justmikejust.wordpress.com/}.
\end{IEEEbiography}
\begin{IEEEbiography}[{\includegraphics[width=.8in,height=1.2in,clip,keepaspectratio]{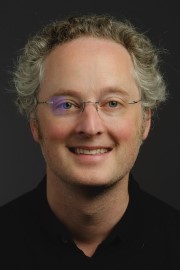}}]{Michael A. Lones}
 (M’01—SM’10) is an Associate Professor of Computer Science at Heriot-Watt University. He received both MEng and PhD degrees from the University of York. He carries out research in the areas of machine learning and optimisation, where he has a particular interest in biologically-inspired approaches. Application areas of his work include medicine, robotics and security. You can find more information at \href{http://www.macs.hw.ac.uk/\%7Eml355}{http://www.macs.hw.ac.uk/$\sim$ml355}.
\end{IEEEbiography}

\captionsetup[table]{name=SM-TABLE}
\captionsetup[algorithm]{name=SM-Algorithm}

\onecolumn
\beginsupplement
\section{Supplementary Material\\ IoTDevID: A behavior-Based Device Identification Method for the IoT}

\maketitle

\begin{table}[htbp]
		\vspace*{13mm}
	\scriptsize
	\centering
	
  \centering
\captionsetup{justification=centering}\caption{Device MAC addresses and labels in Aalto University dataset.\\ MAC addresses in bold correspond to more than one label, and labels in bold correspond to more than one MAC address.}

	\setlength{\tabcolsep}{3pt}
	\begin{tabular}{@{}ll|ll@{}}
		\toprule
		MAC Address & Label & MAC Address & Label \\
		\midrule
		94:10:3e:35:01:c1  & \textbf{WeMoSwitch} &3c:49:37:03:17:db   & \textbf{EdnetCam  } \\
		94:10:3e:34:0c:b5  & \textbf{WeMoSwitch } &3c:49:37:03:17:f0   & \textbf{EdnetCam  } \\
		\textbf{1c:5f:2b:aa:fd:4e  } & D-LinkDoorSensor   & \textbf{00:17:88:24:76:ff } & HueBridge  \\
		\textbf{1c:5f:2b:aa:fd:4e  } & D-LinkHomeHub  & \textbf{00:17:88:24:76:ff } & HueSwitch  \\
		94:10:3e:41:c2:05  & \textbf{WeMoInsightSwitch} &74:da:38:80:7a:08   & \textbf{EdimaxCam } \\
		94:10:3e:42:80:69  & \textbf{WeMoInsightSwitch } &74:da:38:80:79:fc   & \textbf{EdimaxCam } \\
		\bottomrule
	\end{tabular}%
	\label{tab:macs}%
\end{table}%

\begin{table}[htbp]
	\vspace*{13mm}
		\caption{Which data in the UNSW (*\href{https://iotanalytics.unsw.edu.au/iottraces}{IEEE TMC 2018}, **\href{https://iotanalytics.unsw.edu.au/attack-data}{ACM SOSR 2019}) dataset were taken from which date range.}
	\centerline{\includegraphics[width=.61\columnwidth]{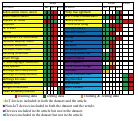}}

	\label{tab:unsw_features}
\end{table}

\setlength{\footskip}{90pt}

\begin{algorithm}[ht]
	\caption{\scriptsize Aggregation Algorithm ($\hat{Y}, M$)} 
	
	\label{alg:c_a}
	\begin{algorithmic}[1]
		\State $MAC_{ij} \leftarrow [\emptyset][\emptyset]$   \Comment{ create an empty two-dimensional array MACs \& index}
		\State $seen\leftarrow [\emptyset]$
		\State $g\leftarrow  12 $ \Comment{specifying the group size} 
		\For  {each $m \in M$} \Comment{discovering unique MAC addresses}
		\If {$m  \notin seen$ }
		\State {$seen \leftarrow seen  \cup m$} 
		\State $i\leftarrow \textbf{length($seen$)}-1,  j \leftarrow 0$ 
		\State {$MAC_{ij}\leftarrow m$ } \Comment{create new array and assign $m$ as the first element}
		\EndIf
		\EndFor

		\For {$j\leftarrow0, $ \textbf{length}$(M)$} \Comment{store the MAC \& indices in the 2D array}
		\For {$i\leftarrow0$, $\textbf{length($seen$)}$ }
		\If {$MAC_{i0} == M_j$ }
		\State {$MAC_i \leftarrow MAC_i\cup j$ }
		\Comment{add the index of $m$  to $MAC_i$}
		
		
		\EndIf
		\EndFor
		\EndFor
		
		\For {$i\leftarrow 0$, $\textbf{length($seen$)}$  }
		\State {$C \leftarrow []$}\Comment{ hold the indices of $m$s  divided into chunks}

		\For {$j\leftarrow 1,$ \textbf{length($MAC_i$)$-g$}, \textbf{Step} $= g$}
		\State $C\leftarrow${\textbf{$C\cup MAC_{i [j:j+g]}$} }
		\Comment{divide  $MAC_i$ into chunks of $g$}
		
		\EndFor
		\For {each $c \in C$}
		\State {$g\_list \leftarrow []$ }
		\For {each $j \in c$}
		\State {$g\_list\leftarrow g\_list \cup \hat{y}_j$} 
		\Comment{{assign $\hat{y}$ sharing same index with m}}
		\EndFor
		\State {$mode \leftarrow$ \textbf{mode($g\_list$) }}

		\For {$j$ \textbf{in $c$}}
		\State {$\hat{y}'_j \leftarrow mode$ } 	\Comment{Aggregated labels}
		\EndFor

		\EndFor
		\EndFor
		
		\State $LABEL_{ij} \leftarrow [\emptyset][\emptyset]$  \Comment{ create an empty 2-dimensional array labels \& MACs}
		\State $seen\leftarrow [\emptyset$]
		\State $exception list\leftarrow [\emptyset] $  
		\State $dominant MACS\leftarrow [\emptyset] $ 
		\For  {each $\hat{y} \in \hat{Y}$}\Comment{discovering unique Labels}
		\If {$\hat{y} \notin seen$ }
		\State {$seen \leftarrow seen  \cup \hat{y}$} 
		\State $i\leftarrow \textbf{length($seen$)}-1,  j \leftarrow 0$ 
		\State {$LABEL_{ij}\leftarrow \hat{y}$ } \Comment{create new array and assign $\hat{y}$ as the first element}
		\EndIf
		\EndFor

		\For {$j\leftarrow0, $ \textbf{length}$(\hat{Y})$}\Comment{store Labels \& MACs in the 2D array}
		\For {$i\leftarrow0$, $\textbf{length($seen$)}$ }
		\If {$LABEL_{i0} == \hat{y}_i$ }
		\State {$LABEL_i \leftarrow LABEL_i\cup m_j$ }
		\Comment{add the index of $m_j$  to $MAC_i$}

		\EndIf
		\EndFor
		\EndFor
		
		\For {$i\leftarrow 0$, $\textbf{length($seen$)}$}	
		\If { \textbf{mode($LABEL_{i}$)} \textbf{not in} $dominant MACS$ }
		\State{	$dominant MACS \leftarrow dominant MACS~\cup$ \textbf{mode($LABEL_{i}$)}}
		\Else
		\State{	$exception list \leftarrow exception list~\cup$ \textbf{mode($LABEL_{i}$)}}
		\EndIf
		\EndFor

		\State $Final Reulsts\leftarrow [\emptyset] $ 
		\For {$j\leftarrow0, $ \textbf{length}$(\hat{Y})$}\Comment{crate mixed labels}
		\If { $m_{j}$ \textbf{in} $exception list$ }
		\State {$FinalResulsts \leftarrow FinalResulsts\cup \hat{y}_j$ }
		\Else
		\State{$FinalResulsts \leftarrow FinalResulsts\cup \hat{y'}_j$}

		\EndIf
		\EndFor
		\State{\textbf{return} $FinalResulsts$}

	\end{algorithmic} 
	
\end{algorithm}

\end{document}